\documentstyle[]{elsart}

\begin{document}
\begin{frontmatter}
\title{Jets from Radio Pulsars}
\author{Jon Bell}
\address{University of Manchester, NRAL, Jodrell Bank, Cheshire,
SK11 9DL, UK. \\ Email: jb@jb.man.ac.uk}

\begin{abstract}
The observational evidence for jets and phenomena arising from rotation
powered radio pulsars is reviewed, including many recent and exciting
discoveries at X-ray wavelengths. The well studied jets of the Crab pulsar
are summarised, including recent results from the HST. The evolutionary
links between the known binary radio pulsars and jets sources in X-ray
binaries are discussed.
\end{abstract}
\end{frontmatter}

\section{Introduction}
In 1986, Feigelson \cite{fei86} reviewed the evidence for jets from neutron
stars and concluded that ``{\it single neutron stars do not make jets but
neutron stars in binary systems can}'', with the speculative corollary that
``the presence of accretion disks'' is responsible for jets in binary
neutron stars. Ten years of further study, new instruments and ideas have
resulted in the discovery of several Galactic systems with jets or
candidates for jets, some of which are associated with single neutron
stars. Accretion powered X-ray binaries are discussed in considerable detail
by Fender and others in these proceedings. In this paper the observational
evidence for jets from rotation powered radio pulsars is reviewed and is
summarised in Table \ref{t:tab}.  The sample of radio pulsars which have
jets or candidate jets is described in \S\ref{s:gen} while the Crab, being
the best example is discussed separately in more detail in \S\ref{s:crab}.
The evolutionary relationships between binary radio pulsars and Galactic jet
sources are highlighted in \S\ref{s:pap}.

\renewcommand\baselinestretch{0.75}
\begin{table}
\begin{center}
\caption{Radio pulsars with jets and related objects. \dag See 
\S\protect\ref{s:crab}}
\label{t:tab}
\begin{tabular}{lll}\hline \hline

Pulsar       & Observational Evidence     & Ref. \\ \hline
B0355+54     & X-ray jet, non-thermal (Einstein) & \cite{hel84a} \\
J0437$-$4715 & H${\alpha}$ bow shock, ram pressure balance  & \cite{bbm+95} \\
             & X-ray nebula (ASCA) & \cite{kaw96} \\
B0531+21\dag & Double X-ray jet + torus (ROSAT) & \cite{asc92,hss+95} \\
             & Single optical jet, line free continuum & \cite{hss+95} \\
             & Perpendicular expansion of counter jet channel & \cite{hss+95} \\
             & Single optical jet, III\,aJ, O{\scriptsize III}, H${\alpha}$, N{\scriptsize II} & \cite{van70,wol57,gf82} \\
             & Perpendicular expansion 260 km\,s$^{-1}$ &\cite{mvw+90,scsn84} \\
             & Linear expansion of $2500$ km\,s$^{-1}$ & \cite{fs93,fg86} \\
             & Radio jet coincident with optical jet & \cite{vel84,fkcg95} \\
             & B field + jet aligned, non-thermal &\cite{wsh85,bk90} \\
B0540$-$69   & H${\alpha}$ jet and N{\scriptsize II} ring  (HST) & \cite{car96} \\
J0633+1746   & X-ray filamentary extended emission (ASCA) & \cite{kaw96} \\
B0656+14     & X-ray jet (ASCA) & \cite{kt96} \\
B0833$-$45   & Wisps interpreted as jet termination shock & \cite{bfh91} \\
             & X-ray jet, non-thermal (ASCA)  & \cite{kaw96} \\
             & X-ray jet (ROSAT, Einstein)  & \cite{mo95,hgs85} \\
             & X-ray bullets (ROSAT)  & \cite{aet95,sjva95}\\
B1046$-$58   & X-ray nebula (ASCA) & \cite{kt96} \\
B1055$-$52   & X-ray jet, non-thermal (Einstein) & \cite{hel84a,kaw96} \\
B1259$-$63   & Radio continuum jet inferred - not resolved & \cite{joh96} \\
             & X-rays from interacting winds (ASCA) & \cite{khnt96} \\
B1509$-$58   & X-ray jet, non-thermal + thermal nebula (ASCA) & \cite{tkyb96} \\
             & X-ray nebula (Einstein) & \cite{sh82} \\
             & X-ray jet (ROSAT) & \cite{gcm+95} \\
B1610$-$50   & X-ray nebula, thermal (ASCA) & \cite{kt96} \\
             & SNR with radio jet & \cite{rmk+85} \\ 
B1718$-$19   & Eclipsing pulsar, illuminated companion & \cite{lbhb93} \\
B1744$-$24A  & Eclipsing pulsar, illuminated companion & \cite{lmd+90} \\
B1757$-$24   & Pulsar out running SNR & \cite{fk91} \\
B1853+01     & Radio bow shock, ram pressure balance & \cite{fggd96} \\
B1929+10     & X-ray nebula (ROSAT) & \cite{yhh94} \\
             & X-ray nebula, non-thermal (ASCA) & \cite{kt96} \\
B1951+32     & H${\alpha}$ bow shock, ram pressure balance & \cite{hk88} \\
             & X-ray nebula (ASCA) & \cite{kt96,sof95} \\
B1957+20     & X-ray nebula (ROSAT) & \cite{fbgb92} \\
             & H${\alpha}$ bow shock, ram pressure balance & \cite{kh88,arc92} \\
             & Eclipsing pulsar, illuminated companion & \cite{aft94} \\
J2051$-$0827 & Eclipsing pulsar, illuminated companion & \cite{sbl+96} \\
B2224+65     & H${\alpha}$ bow shock, ram pressure balance & \cite{crl93} \\ \hline
\end{tabular}
\end{center}
\end{table}
\renewcommand\baselinestretch{1.0}
\normalsize

The term ``jet'' has been applied to detections (at any wavelength) of 
elongated emission regions which may be due to a collimated flows of particles
or material. Although a more rigorous definition, requiring the detection of
proper motion of the jet is desirable there are numerous objects for which
the detection of proper motion is improbable, if not impossible because of
instrumental limitations. For example, it is unlikely that proper motions of
X-ray jets will be measured in the near future.

There are a number of objects associated with pulsars which result from the
emission of relativistic particles which are not collimated. These are
briefly mentioned in this paragraph, with references to recent
reviews. Pulsar wind nebulae result from bow shocks where pulsar winds are
balanced by ram pressure in the interstellar medium (ISM)
\cite{cor96}. While these objects are detected in H${\alpha}$ and are rare,
they may be used to determine pulsar distances, radial velocities, ISM
neutral hydrogen content and the soft X-ray composition of the pulsar
winds. Recently many have been observed in X-rays using ASCA
\cite{kaw96}. There are presently four eclipsing pulsars known and all of
these are millisecond pulsars \cite{fru96}. A significant fraction of the
pulsars' flux at all wavelengths impinges on the companion and heats it. For
two of the four systems this is clearly observed as an optical brightening
of the companion at superior conjunction, and offers constraints on the
composition of the pulsar wind. Finally, there is the case of bullets around
the Vela supernova remnant \cite{aet95,sjva95}. These are detected in both
X-rays and radio and appear to be lumps of material ejected at the time of
the supernova explosion.

\section{Pulsars with Jets}
\label{s:gen}

There are a number of pulsars which may be responsible for jets, though none
of these jets are known to be relativistic.  A quick inspection of
Table~\ref{t:tab} reveals that at present the X-ray band offers the most
prospects, with claims of X-ray jets for at least 6 pulsars. In some cases
(PSRs B0355+54, B1055$-$52, B1929+10) these linear X-ray sources can be
interpreted as mini-Crab nebulae, resulting from non-collimated relativistic
particles confined by ram pressure as the pulsar moves through the ISM,
leaving a wake of X-ray emission
\cite{fei86,hel84a,yhh94}.

There are several cases in which this interpretation does not fit with the
observations and the case for collimated emission is stronger. For the Vela
pulsar (B0833$-$45) the X-ray jet discovered using the Einstein observatory
\cite{hgs85} does not line up with the pulsar proper motion
\cite{bmk+89,car96}. Using a ROSAT image Markwardt and
\"{O}gelman \cite{mo95} incorrectly interpreted the jet as thermal,
while recent ASCA observations clearly revealed a non-thermal spectrum
\cite{kaw96}. There are strongly polarised linear features in the radio
maps of Milne \cite{mil95} that may be associated with the X-ray jet, though
these have not been interpreted as a radio jet. Tamura et al. \cite{tkyb96}
recently used ASCA observations of PSR~B1509$-$52 to demonstrate that it has
both a non-thermal X-ray jet generated by the pulsar and a thermal X-ray
nebula that is created at the working surface of the jet with the ISM.

Apart from the Crab pulsar (see \S\ref{s:crab}), there is little evidence
for jets from radio pulsars at other wavelengths, despite some concerted
searches for them. In particular, at radio wavelengths the regions around
numerous pulsars have been mapped using the VLA \cite{ccgm83}, with no
convincing for jets, except for PSR~B1610$-$50 in the supernova remnant
(SNR) Kes 32 \cite{rmk+85}. Maps at 843 MHz show a well collimated (though
thermal, non-polarised) jet emerging from the SNR and then spreading into a
wide plume. Only the main part of the remnant is detected at X-ray
wavelengths \cite{kaw96}. Many deep optical images have been obtained in
order to study emission from the pulsars or their companions. The only
pulsar (other than the Crab) with a candidate optical jet is PSR~B0540$-$69
\cite{car96}. This was obtained with HST and is seen as a weak double sided
linear feature lying across the pulsar in an H${\alpha}$ light.

\section{Crab Pulsar and Nebula}
\label{s:crab}

With no fewer than {\it three} jets the Crab pulsar must be considered to be
as bizarre as the other Galactic jet sources such at SS433 and
GRO~J1655$-$40. The three jets will be referred to as southeast, northwest
and north.

A ROSAT image of the inner Crab nebula \cite{asc92}, confirmed the presence
of the torus \cite{bal85} and revealed a bright jet southeast of the
pulsar. This jet appears to be aligned with the spin axis and is
perpendicular to the plane of the torus \cite{hss+95} indicating the
presence of collimated relativistic particles from the poles. With
increasing distance from the pulsar, it becomes wider and bends
southward. Wide field ground based images reveal a very similar but slightly
more extended structure in line free optical continuum light. High
resolution HST images in a similar band reveal two knots at 1,400\,AU and
10,000\,AU from the pulsar due to shocks in the inner jet
\cite{hss+95}. Future images of a similar quality and resolution should
allow the measurement of the proper motion of the knots, giving some
indication of the velocity of the jet. The VLA map of Bietenholz and
Kronberg \cite{bk90} shows no evidence for a corresponding radio jet.

A similar but less well defined X-ray  structure in the opposite (northwest)
direction is interpreted as material compressed by the counter jet. In the
optical images there are several features which run parallel to the edges of
the counter jet. From a comparison of images taken at different epochs these
have been shown to have proper motions perpendicularly away from the jet
\cite{hss+95}. Again in the radio maps, there is no evidence for a
radio counter jet \cite{bk90}. 

A third unrelated, jet in a northerly direction, associated with the Crab
nebula was found in a deep optical III\,aJ plate \cite{van70}, although it
was also present on isophote drawings \cite{wol57}. The jet was detected in
the emission lines O{\scriptsize III}, H${\alpha}$ and N{\scriptsize II},
indicating that it has a non-thermal origin \cite{gf82}. It is clearly
present in radio maps made with the VLA \cite{vel84,fkcg95}. The jet is
highly polarised and non-thermal with the local magnetic field lines running
parallel to the jet \cite{wsh85,bk90}. Recent proper motion measurements by
Fesen and Staker \cite{fs93} and Marcelin et al. \cite{mvw+90}
are consistent with less precise earlier results and show that the jet is
expanding along its length at 2500 km\,s$^{-1}$ and perpendicular to its
length at 260 km\,$^{-1}$. Using these velocities and the present jet size
suggest that it formed around the same time as the pulsar. Unlike the other
jets, this jet does not appear to be replenished by the pulsar.

\section{Progenitors and Products}
\label{s:pap}

The discovery of the first millisecond pulsar B1937+21 \cite{bkh+82},
immediately led to a range of formation models (see Bhattacharya and van den
Heuvel, \cite{bv91} for a review). Low mass X-ray binaries (LMXBs) were
among the proposed progenitors and their high incidence in globular clusters
provoked several very successful searches of globular clusters for
millisecond pulsars \cite{lyn92,jkp92}. This provides an excellent example
of how careful consideration of the evolution of a class of objects can
yield methods of finding many more. The progenitors of these LMXBs are
thought to be pulsars with low mass main sequence stellar companions, though
rather surprisingly no such objects have been discovered. Since several
candidates for relativistic Galactic jet sources are associated with LMXBs,
pulsars are therefore potentially both their progenitors and products.

For high mass binary systems, a similar evolutionary picture has emerged in
which a binary system containing a neutron star and a mass main sequence
results from the first supernova explosion \cite{bv91}. While the companion
is still on the main sequence, the neutron star may be visible as a pulsar
and two such objects are known, PSRs B1259$-$63 \cite{jml+92} and
J0045$-$7319 \cite{kjb+94}. When the companion evolves, these systems may
form high mass X-ray binaries (HMXBs), some of which are also thought to
harbour relativistic jets. When the second supernova occurs, either two
solitary neutron stars are formed, or a more exciting dual neutron star
binary in a relativistic orbit is formed, which can be used for tests of
relativity \cite{tw89}. Another possibility is the formation of a neutron
star black hole binary, although no such objects have been found to date.

Lyne and Lorimer \cite{ll94} recently demonstrated that the mean
birth velocity of radio pulsars is 450\,km\,s$^{-1}$. The rotational axis of
the B star in the PSR~J0045$-$7319 system has recently been shown to be
inclined to the orbital plane, giving rise to spin-orbit coupling and
precession of the eccentric pulsar orbit \cite{kbm+96}. This has provided
independent evidence for a substantial neutron star birth kick. In general
therefore, HMXBs, including those with jets might also be expected to have
their orbital and spin angular momenta misaligned. At its last periastron
passage in January 1994, PSR~B1259$-$63 revealed some remarkable
interactions with its Be star companion, including the disappearance of the
pulsar, huge dispersion measure and rotation measure variations, spin-up of
the pulsar due to accretion and a continuum source possibly due to a wind
interaction or jet \cite{jml+96,joh96}. Recent X-ray observations
are more easily explained by the wind interaction model \cite{kas96}.

\section{Conclusions}

There appears to be significant evidence that isolated pulsars or neutron
stars can produce jets, though none of these appear to be relativistic.
There are many exciting prospects, particularly in the X-ray band. The
existence of 3 jets associated with the Crab pulsar and nebula seems to be
well established. Two of these jets are clearly associated with the pulsar
and lie along its spin axis. Binary radio pulsars are likely to be both
progenitors and products of Galactic jet sources in X-ray binaries and all of
these systems are likely to have their  spin and orbital angular momenta
misaligned.

\renewcommand\baselinestretch{0.2}


\renewcommand\baselinestretch{1.0}

\end{document}